\documentstyle[12pt]{article}
\begin{document}
\large

\begin{center}
{\Large \bf Linear determining equations, differential constraints
and invariant solutions.}
\end{center}

\begin{center}
{\bf Oleg V Kaptsov and Alexey V Schmidt}

 Institute of Computing Modelling RAS, Akademgorodok,\\
  660036, Krasnoyarsk, Russia

   E-Mail: kaptsov@ksc.krasn.ru
\end{center}

{\bf Abstract}

A construction of differential constraints compatible with partial
differential equations is considered. Certain linear determining
equations with parameters are used to find such differential
constraints. They generalize the classical determining equations
used in the search for admissible Lie operators. As applications
of this approach non-linear heat equations and Gibbons-Tsarev's
equation are discussed. We introduce the notion of an invariant
solution under an involutive distribution and give sufficient
conditions for existence of such a solution.

\vspace{05mm}


\vspace{15mm}
 {\LARGE 1. Introduction.}

As is well known, one can produce many exact solutions of partial
differential equations by means of additional constraints
\cite{gou1}, \cite{sid2}. Differential constraints arisen
originally in the theory of partial differential equations of the
first order. Lagrange in particular used differential constraints
to find the total integral of nonlinear equation
 $$ F(x,y,u,u_x,u_y) = 0 . $$
Darboux \cite{dar3} applied differential constraints to integrate
the partial differential equations of the second order. The
detailed description of the Darboux method can be found in
\cite{gou1}, \cite{for}.

The general formulation of the method of differential constraints
requires that  the original system of partial differential
equations
 $$ F^1 = 0,..., \quad F^m = 0      \eqno (1.1) $$
 be enlarged by appending additional differential equations (differential
constraints)
 $$ h_1 = 0,..., \quad h_p = 0,  \eqno (1.2) $$
 such that the over-determined system (1.1), (1.2) satisfies some conditions
of compatibility.

The theory of over-determined systems was developed by Delassus,
Riquier, Cartan, Ritt, Kuranishi, Spencer and others. One can find
references in the book of Pommaret  \cite{pom}. Now the
applications of over-determined systems include such diverse
fields as differential geometry, continuum mechanics and nonlinear
optics. Unfortunately the problem of finding all differential
constraints compatible with certain equations can be more
complicated than the investigation of the original equations.
Therefore it is better to content oneself with finding constraints
in some classes, and these classes must be chosen using additional
considerations.

It was recently proposed a new method for finding differential
constraints which uses linear determining equations. These
equations are more general than the classical determining
equations for Lie generators \cite{ovs} and depend on some
parameters.
 Given an evolution equation
$$ u_t = F(t,x,u,u_1,\dots,u_n) ,   \eqno (1.3) $$
 where $u_k = \frac{\partial^k u}{\partial x^k}$,  then according
to \cite{kap1} the linear determining equation corresponding to
(1.3) is of the form
 $$D_t(h) = \sum^{n}_{i=0} \sum^{i}_{k=0} b_{ik}
D^{i-k}_x(F_{u_{n-k}}) D^{n-i}_x(h) ,\qquad
b_{ik}\in R.\eqno(1.4)$$
 Here and throughout $D_t, D_x$ are the operators of
total differentiation with respect to $t$ and $x$. Equality (1.4)
must hold for all solutions of (1.3). The function $h$ may depends
on $t, x, u, u_1,\dots,u_p$. The number $p$ is called the order of
the solution of the equation (1.4).
 If we have some solution $h$, the corresponding differential constraint is
$$ h =0.   \eqno(1.5) $$
 It was also shown in \cite{kap1} that the equations
(1.4) and (1.5) constitute the compatible system.

The organization of this chapter as follows. In section 2 we
introduce the requisite concepts and derive a nonlinear
determining equation. Linearizing and slightly this equation
modifying one we obtain the linear determining equation. In
section 3 we focus on the solutions of the second and third orders
to the linear determining equation  for Gibbons-Tsarev's equation
\cite{gib1}
 $$ u_{tt} = u_x u_{tx} - u_t u_{xx} + 1 . \eqno (1.6)$$
This gives the corresponding differential constraints and allow us
to find some exact solutions of (1.6). Section 4 is devoted to
systems of reaction-diffusion equations
 $$ u_t = (u^kv^lu_x)_x + f_1(u,v),$$
 $$ v_t = d_1(u^mv^nv_x)_x + f_2(u,v) .$$
In the final section the invariant solutions under an involutive
distribution are discussed. We consider the problem of finding
involutive distributions that enable us to obtain invariant
solutions to evolution equations.

\vspace{15mm}
 {\LARGE  2. Linear determining equations for differential
constraints.}

We consider a system $S$ of evolution equations:
 $$   u_t^i = F^i(t,x,u^1,\dots,u^m,u^1_x,u^2_x,\dots ), \quad i=1,\dots,m.
\eqno(2.1) $$
 The right-hand sides of (2.1) can depend on the derivatives of
arbitrary orders with respect to $x$. For simplicity we assume
that the $F^i$ are smooth functions. We supplement $S$ by
differential constraints $H$:
 $$ h_j(t,x,u^1,\dots,u^m,u^1_x,u^2_x,\dots ) = 0,
\quad j=1,\dots,p \eqno(2.2)$$ where $p\leq m$. The functions
$h_i$ can depends on derivatives of arbitrary orders with respect
to $x$, but contain no $t$-derivatives.

According to \cite{kap2}, the differential constraints $H$ define
an invariant manifold of the system $S$ if
 $$ D_t (h_j)_{\mid[S]\cap[H]} = 0 , \quad j=1,\dots,p . \eqno(2.3)$$
  We denote by $[S]$ the equations (2.1) and their differential
consequences with respect to $x$. The constraint (2.2) and its
differential consequences with respect to $x$ are denoted by
$[H]$.
 %We denote the operators of total differentiation with respect to
%$t$ and $x$ by $D_t$ and $D_x$ respectively.

If there are $m$ differential constraints which are all resolved
with respect to highest derivatives of all functions
$u^1,...,u^m$, then, as shown in \cite{kap3} the invariance
condition (2.3) yields the existence of a smooth solution of the
system (2.1), (2.2).

For simplicity we now consider an equation of the second order
 $$ u_t = F(t,x,u,u_1,u_2)  \eqno (2.4) $$
 and the differential constrain
 $$ h(t,x,u,u_1,\dots,u_n) = 0, $$
 where $u_k = \frac{\partial^k u}{\partial x^k}$. We denote by
 $[E]$ the equation (2.4) and its differential consequences.

The invariance condition (2.3) is equivalent to the following
nonlinear equation

 $$D_t(h) \mid_{[E]} = F_{u_2}D^2_x(h) + \bigg(F_{u_1} + nD_x(F_{u_2})\bigg)D_x(h)
 +$$
 $$+ \Bigg(F_u +
 nD_x(F_{u_1}) -  h_{u_{n-1}}D_x(F_{u_2}) +
\frac{n(n-1)}{2}D^2_x(F_{u_2}) +$$ $$+  F_{u_2}hh_{u_{n-1}u_{n-1}}
-  2F_{u_2}D_x(h_{u_{n-1}})\bigg)h,  \eqno(2.5) $$
 with $n\ge 4$.

 Indeed, it is easy to see that
 $$ D_t (h)_{\mid [E]} \simeq D^n_x(F) + h_{u_{n-1}}D^{n-1}_x(F) +
h_{u_{n-2}}D^{n-2}_x(F) . \eqno(2.6)  $$
 Here and throughout we
write  $\alpha \simeq \beta$ to indicate that there are no terms
including $u_n, u_{n+1}, u_{n+2}$ in the difference $\alpha -
\beta$ . Since  $n\ge 4$, the terms on the right-hand side of
(2.6) can be represented as follows:
 $$D^n_x(F) \simeq F_{u_2} u_{n+2} + [F_{u_1} +
nD_x(F_{u_2})]u_{n+1} + \Bigg[ F_u + nD_x(F_{u_1})  +
\frac{n(n-1)}{2}D^2_x(F_{u_2})\Bigg]u_n,$$
 $$h_{u_{n-1}}D^{n-1}_x(F) \simeq h_{u_{n-1}}\Bigg[u_n\bigg(F_{u_1} + (n
- 1)D_x(F_{u_2})\bigg) + u_{n+1}F_{u_2}\Bigg],$$
 $$h_{u_{n-2}}D^{n-2}_x(F) \simeq u_n F_{u_2} h_{u_{n-2}}.$$

 Hence (2.6) can be written as the relation
 $$ D_t (h)_{\mid [E]} \simeq F_{u_2}
u_{n+2} + u_{n+1}[F_{u_1} + nD_x(F_{u_2}) + F_{u_2}h_{u_{n-1}}] +
\Bigg[F_u +$$ $$+ \frac{n(n-1)}{2}D^2_x(F_{u_2}) +
 nD_x(F_{u_1}) +
h_{u_{n-1}}(F_{u_1} + nD_x(F_{u_2})) + F_{u_2} h_{u_{n-2}}\Bigg]u_n .$$
 It is easy to see that
 $$D_x(h)\simeq u_{n+1} + u_n h_{u_{n-1}},$$
 $$D^2_x(h) \simeq  u_{n+2} +
u_{n+1} h_{u_{n-1}} + u_n[h_{u_{n-2}} + 2D_x(h_{u_{n-1}}) - u_n
h_{u_{n-1}u_{n-1}}].$$
 Hence the difference
 $$ D_t (h)_{\mid [E]} - F_{u_2}D^2_x(h) - [F_{u_1} +
nD_x(F_{u_2})]D_x(h)  $$
 contains no terms with $u_{n+2}$ and $u_{n+1}$.
 A direct calculation shows that there are no terms containing $u_n$
in the expression for the function
  $$\gamma = D_t(h)_{\mid [E]} - M(h) ,$$
 where
 $$M(h) = F_{u_2}D^2_x(h) + [F_{u_1} + nD_x(F_{u_2})]D_x(h) +
  \Bigg[F_u + nD_x(F_{u_1}) - $$
$$- h_{u_{n-1}}D_x(F_{u_2}) +
 \frac{n(n-1)}{2}D^2_x(F_{u_2}) +  F_{u_2}hh_{u_{n-1}u_{n-1}} -
  2F_{u_2}D_x(h_{u_{n-1}})\Bigg]h ,$$
that is, $\gamma \simeq 0$. We claim that $\gamma$ is equal to
zero. Since $H$ is an invariant manifold, it follows that
 $$M(h) + \gamma = 0 $$
 on the set $[E]\cap[H]$. Clearly, $M(h)$ vanish there, therefore
 $$\gamma_{\mid [E]\cap[H]} = 0.$$
 Since $\gamma$ is independent of $u_t$, $u_{tx},\dots$, we can
rewrite the last equality as follows:
 $$\gamma_{\mid [H]} = 0. $$
 As shown above, $\gamma \simeq 0$, that is $\gamma$ can depend
only on $u_{n-1}, u_{n-2},\dots .$ On the other hand, $h$ depends
on $u_n$. Hence  $\gamma$ is equal to zero.

It is clear that the equation (2.5) is difficult to solve,
therefore, in place of the non-linear equation, we propose to use
in the search for invariant manifolds linear equations of a
similar kind. This leads to the following definition.

{\it Definition.} The equation of the form
 $$D_t(h)_{\mid [E]} = F_{u_2}D^2_x(h) + \bigg(c_1 F_{u_1} + c_2
D_x(F_{u_2})\bigg)D_x(h) + $$
 $$ \Bigg(c_3 F_u + c_4 D_x(F_{u_1})
+ c_5 D^2_x(F_{u_2}) \bigg)h  \eqno(2.7) $$
 is called the linear determining equation corresponding to (2.4).
Here $c_1,\dots,c_5$ are some constants.

It is easy to generalize the previous definition for case of
equation (1.3).

{\it Definition.} A linear determining equation corresponding to
(1.3) is the equation (1.4).

 The nonlinear diffusion equation
 $$u_t = (u^k u_x)_x + f(u)  $$
 have been considered in \cite{Kap-Ver}. It is shown that the
solutions of linear determining equations of the second and third
orders exist only if $f$ belongs to the special forms.

In particular the equation
 $$u_t = (u^{-1/2}u_x)_x + m u -2k\sqrt u , \qquad m,k \in R$$ %\eqno (2.8)$$
 is compatible with differential constrain
 $$  u_3 - \frac{5u_1u_2}{2u} + \frac{5u_1^3}{4u^2} +
r e^{-3mt/2} u^{5/2}  + s e^{mt/2}\sqrt u = 0 , \qquad r,s \in R.$$%\eqno(33) $$
 This leads to the following representation
  $$ u = -\frac{ 2 X^{\prime}T^{\prime} }{(X+T)^2} e^{3mt/2} . $$
The functions $T(t)$ and $X(x)$ satisfy the ordinary differential
equations
 $$(X^{\prime})^3 = ( c_3 X^3 +c_2 X^2 +c_1 X +c_0 )^2
 ,$$ %\eqno (35)$$
 $$(T^{\prime})^3 = A ( -c_3 T^3 +c_2 T^2 -c_1 X +c_0 )^2 ,$$ %\eqno(36) $$
where $c_3, c_2, c_1$ and $c_0$ are arbitrary constants, $A=
(-2r)^{1/3}$. It is possible to show that the functions $X$ and
$T$ can be expressed by means of elliptic functions
\cite{Kap-Ver}.
 The others examples can be found in \cite{Kap-Ver}.

The linear determining equation corresponding to the equation
 $$u_{tt} = F(t,x,u,u_1,\dots,u_n) \eqno(2.8)$$
 is as follows:
 $$D_{tt}(h)|_{[E]} = \sum^{n}_{i=0} \sum^{i}_{k=0} b_{ik}
D^{i-k}_x(F_{u_{n-k}}) D^{n-i}_x(h) ,\qquad b_{ik} \in
R,\eqno(2.9) $$
 where $[E]$ means the equation (2.8) and its differential
consequences.

{\it Example.} Let us consider the equation
 $$ u_{tt} = u_{xx} + \frac{a}{t}u_t + \frac{b}{x}u_x + \sin(u), \eqno(2.10)$$
It is easy to check that the symmetry group \cite{ovs} for (2.1)
is trivial. On the other hand, the linear determining equation
 $$D_{tt}h = D_{xx}h + \frac{a}{t}D_t h + \frac{b}{x}D_x h +
 (\cos(u) - \frac{a}{t^2} - \frac{b}{x^2})h = 0 $$
corresponding to (2.10) has the solution  $ h = xu_t + tu_x .$
 Differential constraint
 $$ h = xu_t + tu_x = 0$$
 gives the following representation
 $$ u = V(x^2 - t^2) \eqno (2.11)$$
 for a solution of (2.10). Substituting (2.11) into (2.10) we have
 the ordinary differential equation
 $$ 4zV^{\prime\prime} + 2(b-a+2)V^{\prime} + \sin(V) = 0. $$

\vspace{15mm}
 {\LARGE 3. Gibbons-Tsarev's equation.}

In this section we will consider the Gibbons-Tsarev equation
\cite{gib1}
 $$z_{xx}+z_yz_{xy}-z_xz_{yy}+1=0.\eqno(3.1)$$
 which arises in reductions of the Benney equation.

By (2.9), the linear determining equation has the form
$$D_{x}^2h+z_yD_xD_yh-z_xD_y^2h+b_1z_{yy}D_xh+b_2z_{xy}D_yh=0,\eqno{(3.2)}$$
The constants $b_1$ and $b_2$ are to be determined together with
the function $h$.

It can be shown that the equation (3.2) has a solution of the form
  $$h=z_{yy}+g(x,y,z,z_x,z_y,z_{xx})$$
 if and only if the function $g$ is independent of $z_{xx}$.
 Therefore we shall start with solutions of the second order
 $$h=z_{yy}+g(x,y,z,z_x,z_y).\eqno{(3.3)}$$
 Substituting (3.3) into (3.2) leads to an equation which includes
derivatives of the third order. We can express all mixed
derivatives by means of (3.1). Setting the coefficients of
$z_{xxx}$ and $z_{yyy}$ equal to zero we obtain :
  $$b_1 = 1, \qquad b_2 = -1 .$$
 The left-hand side of (3.2) is a polynomial with respect to
$z_{xx}$ and $z_{yy}$. This polynomial must identically vanish.
Collecting similar terms we have the following equations:
     $$pg_{pp} + qg_{pq} - g_{qq} + 2g_{p} = 0, \eqno{(3.4)}$$
$$(-q^2-2p)g_{pp} + 2g_{qq} + q^2g_{xp} + q(q^2+2p)g_{yp} +
q^2(q^2+3p)g_{zp} -$$ $$- 2qg_{xq}  - q^2g_{yq} - q(q^2+2p)g_{zq}
 + qg_y + q^2g_z - 4g_p = 0,\eqno{(3.5)}$$
    $$2p^2g_{pp} - (q^2+2p)g_{qq} + pq^2g_{xp} - 2p^2qg_{yp} -
p^2q^2g_{zp} +$$ $$+ q(q^2+2p)g_{xq} - pq^2g_{yq} + 2p^2qg_{zq} +
q^2g_x - pqg_y + 4pg_p = 0, \eqno{(3.6)}$$
 $$ pg_{pp} - g_{pq} -
g_{qq} + q^2g_{xp} - 2pqg_{yp} - pq^2g_{zp} + 2qg_{xq} + q^2g_{yq}
+ q(q^2+2p)g_{zq} -$$
 $$- q^2(q^2+2p)g_{xz} + pq^3g_{yz} - p^2q^2g_{zz} + 2g_{p} - q^2g_{xx}
- q^3g_{xy} + pq^2g_{yy} - qg_y = 0,\eqno{(3.7)}$$
 where $p = z_x$ and $q = z_y$.

It is possible to show that the general solution of the equations
(3.4) -- (3.7) is
  $$h = z_{yy} + c_1(z_y^4+(3z_x+4x)z_y^2+3yz_y+(z_x+2x)^2+2z) +$$
$$+ c_2(z_y^3+(2z_x+3x)z_y+2y) + c_3(z_y^2+z_x+2x)+c_4z_y+c_5. $$
Hence the differential constraint $h = 0$ is compatible with the
Gibbons-Tsarev equation (3.1). In the case $c_1=c_2=c_3=0$
we obtain the differential constraint
  $$z_{yy} + c_4z_y + c_5=0.\eqno(3.8)$$
 From (3.8) we find the following representation
  $$z = a_1 \exp(-c_4y) - c_5y/c_4 + a_2,\eqno(3.9)$$
where $a_1$ and $a_2$ depend on $x$. Substituting (3.9) into (3.1)
we derive two ordinary differential equations
  $$a_2''+1=0, \qquad  a_1''+c_5a_1'-c_4^2a_1a_2'=0.$$

The first equation has the solution
  $$a_2 = - x^2/2 + c_6x + c_7 . \qquad c_6,c_7 \in R$$
In this case the second equation is
  $$a_1''+c_5a_1'+c_4^2(x-c_6)a_1=0. $$
 Setting $a_1 = \exp(-c_5x/2)v(x)$ we obtain the equation
  $$ v^{''} + (A +Bx)v = 0, \qquad A,B\in R.$$
According to \cite{olv2}
the solutions of last equation can be
expressed in terms of the Airy functions.

It can be shown that the linear determining equation (3.2) has the
following solution of the third order %( in this case
 $$h=z_{yyy}+c_1(3z_y^5+(10z_x+12x)z_y^3+6yz_y^2+(6z_x^2+18xz_x+2z+12x^2)z_y+$$
$$+4yz_x+6xy)+c_2(5z_y^4+(12z_x+15x)z_y^2+6yz_y+3z_x^2+10xz_x+15/2x^2+$$
$$+z)+c_3(2z_y^3+(3z_x+4x)z_y+y)+c_4(3z_y^2+2z_x+3x)+c_5z_y+c_6.$$
 The corresponding constants
$b_1$ and $b_2$ in (3.2) are given by
  $$b_1=2 , \qquad b_2=-2 . $$
In the case  $c_1=c_2=c_3=c_6=0$ and $c_5=-1$  the function
$h$ gives the differential constraint
  $$ z_{yyy} - z_y = 0 . \eqno{(3.10)}$$
From (3.10) we obtain the following representation
  $$z = s_{1}(x)+s_{2}(x)e^{y}+s_{3}(x)e^{-y}.$$
The functions $s_{1}(x)$, $s_{2}(x)$ and $s_{3}$ must satisfy the
equations
  $$s_{2}^{''} - s_{1}^{'}s_{2}=0,$$
 $$s_{1}^{''} - 2s_{3}s_{2}^{'} - 2s_{2}s_{3}^{'} + 1=0,$$
 $$s_{3}^{''} - s_{1}^{'}s_{3}=0.$$
 If  $s_3 = as_2$ then the last system reduces to the two equations
 $$s_{2}^{''} - s_{1}^{'}s_{2}=0, \qquad
 s_{1}^{''}-4as_{2}s_{2}^{'} +1=0 , \quad
  a\in R. \eqno (3.11)$$
 Integrating the second equation, we find that
 $$s_1^{'}= - t - b + 2as_{2}^{2}, \qquad  b\in R.$$
 We can insert this expression in (3.11) and obtain
the second-order equation
 $$s_{2}^{''} + (t+b-2as_{2}^{2})s_{2}=0$$
Using the transformations  $t_{1}=t+b$ and
$w=\sqrt{a}s_{2}$, we take the equation in $s_2$ to the second
Painleve equation \cite{ince}
 $$w^{''}=2w^{3}+t_{1}w.$$
The differential constraint
 $$ z_{yyy} = 0 $$
leads to a solution of (3.1) which is expressed in terms of
elementary functions.

\vspace{15mm}
{\LARGE 4. The reaction-diffusion equations.}

In this section we consider systems of the second order

 $$u_t=(u^kv^lu_x)_x + f_1 , \eqno{(4.1)}$$
$$v_t=d_1(u^mv^nv_x)_x + f_2 , \eqno{(4.2)}$$
where $k$, $l$, $m$, $n$, and $d_1$ are arbitrary constants, the functions $f_1$ and $f_2$ depend on $u$ and $v$. This model plays important role in various applications.

Using the method described in the section 2 it is possible to obtain the linear determining equations to the system (4.1) -- (4.2). We give the equations in the final shape:

$$D_t(h)=u^kv^lD_x^2(h)+(b_1ku^{k-1}v^lu_x+b_2lu^kv^{l-1}v_x)D_x(h)+$$
$$+b_3lu^kv^{l-1}u_xD_x(\beta)+\bigg(b_4f_{1u}+(b_4+2b_5+b_6)k(u^{k-1}v^lu_{xx}+
$$
$$+(k-1)u^{k-2}v^lu_x^2)+(b_4+3b_5+b_6)klu^{k-1}v^{l-1}u_xv_x+(b_5+$$
$$+b_6)l(u^kv^{l-1}v_{xx}+(l-1)u^kv^{l-2}v_x^2)\bigg)h+\bigg(b_7f_{1v}+$$
$$+b_8l(u^kv^{l-1}u_{xx}+ku^{k-1}v^{l-1}u_x^2+(l-1)u^kv^{l-2}u_xv_x)\bigg)\beta,           \eqno{(4.3)}$$

$$D_t(\beta)=d_1u^mv^nD_x^2(\beta)+d_1(b_9mu^{m-1}v^nu_x+b_{10}nu^mv^{n-1}v_x)D_
x(\beta)+$$
$$+b_{11}d_1mu^{m-1}v^nv_xD_x(h)+\bigg(b_{12}f_{2u}+b_{13}d_1m(u^{m-1}v^nv_{xx}+
$$
$$+nu^{m-1}v^{n-1}v_x^2+(m-1)u^{m-2}v^nu_xv_x)\bigg)h+\bigg(b_{14}f_{2v}+d_1n(b_
{14}+$$
$$+2b_{15}+b_{16})(u^mv^{n-1}v_{xx}+(n-1)u^mv^{n-2}v_x^2)+d_1(b_{14}+3b_{15}+$$
$$+b_{16})mnu^{m-1}v^{n-1}u_xv_x+d_1m(b_{15}+b_{16})(u^{m-1}v^nvu_{xx}+$$
$$+(m-1)u^{m-2}v^nu_x^2)\bigg)\beta .          \eqno{(4.4)}$$ Here
$h$ and $\beta$ are required functions, $b_{ij}$ are some
constants.

At first we seek for solutions of the second order to (4.3)-(4.4)
of the form $$h=u_{xx}+h_1(t,x,u,v,u_x,v_x),  \
 \beta=v_{xx}+\beta_1(t,x,u,v,u_x,v_x) . $$

We do not itemize the systems (4.1)-(4.2) which lead to the
simplest solutions
 $$h=u_{xx} ,  \   \beta=v_{xx} $$
 of the equations (4.3)-(4.4). Moreover, we did not include
solutions generated by the symmetry groups.

The full list of nonlinear systems and corresponding solutions
 of linear determining equations is as follows:
$$
\begin{array}{ll}
u_t=(uu_x)_x+2f_{11}u+f_{12}v+f_{13},&  h_{1}=u_{xx}+f_{11},\\[1mm]
v_t=d_1v_{xx}+f_{21}u- {\displaystyle \frac{f_{21}f_{12}}{f_{11}}}v+f_{23},
& \beta_{1}=v_{xx}+ {\displaystyle \frac{f^2_{11}}{f_{12}}};\\
 & \\
u_t=(uu_x)_x+2f_{11}u+f_{12}v+f_{13}, &h_{2}=u_{xx}+f_{11},\\[1mm]
v_t\!=\!d_1(uv_x)_x\!+\!f_{21}u\!+\!(3d_1f_{11}\!-\!2f_{12})v\!+\!f_{23},
&
\beta_{2}\!=\!v_{xx}\!+\!{\displaystyle \frac{f^2_{11}}{f_{12}}};\\
 & \\
u_t=(uu_x)_x+{\displaystyle \frac{f_{11}}2}u^2+f_{12}v-
{\displaystyle \frac{f_{21}f_{12}}{f_{11}}},
&
h_{3}=u_{xx}+ {\displaystyle \frac{f_{11}}3u},\\
v_t=-2(uv_x)_x\!+\!f_{21}u\!-\!f_{11}uv\!-\!{\displaystyle \frac{f^2_{11}}{2f_{12}}}u^3,
&
\beta_{3}\!=\!v_{xx}\!+\!{\displaystyle \frac1{3f_{12}}}\Bigg(f_{11}f_{12}v\!+
\!{\displaystyle \frac{f_{11}^2}2}u^2\!-\\
& -\!f_{21}f_{12}\Bigg);\\
 & \\
u_t\!=\!(uu_x)_x\!+\!f_{11}uv\!+\!f_{12}u^2\!+\!f_{13}u\!+\!f_{14},
&
h_{4}=u_{xx}+ {\displaystyle \frac{f_{11}}{3}}v+{\displaystyle
\frac{2f_{12}}{3}}u+ {\displaystyle \frac{f_{13}}3},\\
v_t=\!-2(uv_x)_x\!-
{\displaystyle \frac{2f_{12}}{f_{11}}}(f_{11}uv\!+\!f_{12}u^2+
&
\beta_{4} =0;\\
+ f_{13} u +f_{14}), & \\
 & \\
u_t=(uu_x)_x+f_{11}u+f_1(v),
&
h_{5}=u_{xx}+f_{11},\\[1mm]
v_t=(f_{21}u\!+\!
f_{22}\!-\!{\displaystyle \frac{f_{21}}{2f_{11}}}f_1(v))/f'_1(v),
&
\beta_{5}
\!=\!v_{xx}\!+\!
{\displaystyle \frac{f''_1(v)}{f'_1(v)}}v_x^2\!+\!{\displaystyle \frac{2f^2_{11}}{f'_1(v)}};\\
 & \\
\end{array}
$$
$$
\begin{array}{ll}
u_t=(uu_x)_x+f_{11}u+f_{12}v+f_{13},
&
h_{6}=u_{xx}+f_{11},\\[1mm]
v_t\!=\!d_1(vv_x)_x\!+\!
{\displaystyle \frac{2f_{11}}{f^2_{12}}}
(6d_1f^2_{11}\!-\!f_{12}f_{22})u+
&
\beta_{6}=v_{xx}+{\displaystyle \frac{2f^2_{11}}{f_{12}}};\\
+f_{22}v+f_{23},& \\
 & \\
u_t=u_{xx}+f_{11}u-{\displaystyle \frac{f_{22}f_{11}}{f_{21}}}v+f_{13},
&
h_{7}=u_{xx}+ {\displaystyle \frac{2f_{22}}3},\\[1mm]
v_t=d_1(uv_x)_x+d_1f_{21}u+d_1f_{22}v+f_{23},
&
\!\beta_{7}\!=\!v_{xx}\!+\!{\displaystyle \frac{2f_{21}}3};\\
 & \\
u_t=u_{xx}+f_{11}u+f_{12}v+f_{13},
&
h_{8}\!=\!u_{xx}\!+\!{\displaystyle \frac{f_{12}}{3f^2_{11}d_1}}
(f_{21}\!f_{12}-\!f_{22}\!f_{11}),\\[1mm]
v_t=d_1(vv_x)_x+f_{21}u+f_{22}v+f_{23},
&
\beta_{8}\!=\!v_{xx}\!-\!
{\displaystyle \frac{f_{21}f_{12}\!-\!f_{22}f_{11}}{3f_{11}d_1}};\\
 & \\
u_t=(vu_x)_x+f_{11}u+f_{12}v+f_{13},
&
h_{9}=u_{xx}+{\displaystyle \frac{f_{12}f_{21}-f_{11}
f_{22}}{3f_{21}}},\\[2mm]
v_t=d_1v_{xx}+f_{21}u+f_{22}v+f_{23},
&
\beta_{9}=v_{xx}-{\displaystyle \frac{f_{12}f_{21}-f_{11}f_{22}}{3f_{22}}};\\
 & \\
u_t=(vu_x)_x+f_{11}u+f_{12}v+f_{13},
&
h_{10}\!=\!u_{xx}\!+\!{\displaystyle \frac{f_{11}f_{22}\!-\!f_{12}
f_{21}}{3(d_1f_{11}\!-\!f_{21})}},\\[2mm]
\end{array}
$$
$$
\begin{array}{ll}
v_t=d_1(uv_x)_x+f_{21}u+f_{22}v+f_{23},
&
\beta_{10}\!=\!v_{xx}\!+\!
{\displaystyle \frac{f_{12}f_{21}\!-\!f_{11}f_{22}}{3(d_1f_{12}\!-\!f_{22})}};\\
 & \\
u_t=(vu_x)_x+f_{11}u+f_{12}v+f_{13},
&
h_{11}\!=\!u_{xx}\!-\!
{\displaystyle \frac{f_{12}f_{22}}{d_1f_{11}\!-\!3f_{22}}},\\[2mm]
v_t=d_1(vv_x)_x-{\displaystyle \frac{2f_{22}(d_1f_{11}-3f_{22})}{d_1f_{12}}}u,
&
\beta_{11}=v_{xx}+ {\displaystyle \frac{f_{22}}{d_1}}+f_{22}v+f_{23};\\
 & \\
u_t=(vu_x)_x+f_{11}u+{\displaystyle \frac{5f_{11}f_{22}}{f_{21}}}v+f_{13},
&
h_{12}\!=\!u_{xx}\!+\!{\displaystyle \frac{f_{11}f_{22}}{2f_{21}}},\\[2mm]
v_t=f_{21}u+f_{22}v+f_{23},
&
\!\beta_{12}\!=\!v_{xx}\!\!-\!\!{\displaystyle \frac{f_{11}}2}.\\
\end{array}
$$
$$
\begin{array}{ll}
u_t=(v^lu_x)_x+f_{11}u+{\displaystyle \frac{5f_{11}f_{23}}{2f_{21}}}v^l+f_{13},
&
h_{13}=u_{xx}+{\displaystyle \frac{f_{11}f_{23}}{2f_{21}}},\\
v_t=f_{21}v^{1-l}u+f_{22}v^{1-l}+f_{23}v,
&
\beta_{13}=v_{xx}+(l-1){\displaystyle \frac{v^2_x}{v}-\frac{f_{11}}{2l}
v^{1-l}},\\
 & \\
u_t=(uu_x)_x+f_{11}u+f_{12}v^{1/4}+f_{13},
&
h_{14}=u_{xx},\\
v_t=d_1(v^{1/2}v_x)_x+f_{21}uv^{3/4}+f_{22}v^{3/4}+f_{23}v,
&
\beta_{14}=v_{xx}-{\displaystyle \frac{3v_x^2}{4v}};\\
 & \\
u_t=(uu_x)_x+f_{11}u+f_{12}v^{3/2}+f_{13},
&
h_{15}=u_{xx},\\
v_t=d_1(v^{1/2}v_x)_x+f_{21}uv^{-1/2}+f_{22}v^{-1/2}+f_{23}v,
&\beta_{15}=v_{xx}+  {\displaystyle \frac{v_x^2}{2v}};\\
 & \\
u_t=(u^{1/2}u_x)_x+f_{11}u^{3/4}v+f_{12}u^{3/4}+f_{13}u,
&
h_{16}=u_{xx}-{\displaystyle \frac{3u_x^2}{4u}},\\
v_t=d_1(vv_x)_x+f_{21}v+f_{22}u^{1/4}+f_{23},
&
\beta_{16}=v_{xx};\\
 & \\
u_t=(u^{1/2}u_x)_x+f_{11}u^{-1/2}v+f_{12}u^{-1/2}+f_{13}u,
&
h_{17}=u_{xx}+{\displaystyle \frac{u_x^2}{2u}},\\
v_t=d_1(vv_x)_x+f_{21}v+f_{22}u^{3/2}+f_{23},
&
\beta_{17}=v_{xx};\\
 & \\
u_t=u_{xx}+f_{11}u+f_{12}v^{1/4}+f_{13},
&
h_{18}=u_{xx},\\
v_t=d_1(v^{1/2}v_x)_x+f_{21}uv^{3/4}+f_{22}v^{3/4}+f_{23}v,
&\beta_{18}=v_{xx}-{\displaystyle \frac{3v_x^2}{4v}};\\
 & \\
u_t=u_{xx}+f_{11}u+f_{12}v^{3/2}+f_{13},
&
h_{19}=u_{xx},\\
v_t=d_1(v^{1/2}v_x)_x+f_{21}uv^{-1/2}+f_{22}v^{-1/2}+f_{23}v,
&
\beta_{19}=v_{xx}+{\displaystyle \frac{v_x^2}{2v}};\\
 & \\
u_t=(u^{1/2}u_x)_x+f_{11}u^{3/4}v+f_{12}u^{3/4}+f_{13}u,
&
\!h_{20}\!=\!u_{xx}\!-\!{\displaystyle \frac{3u^2_x}{4u}},\\
v_t\!=\!d_1v_{xx}\!+\!f_{21}v\!+\!f_{22}u^{1/4}\!+\!f_{23},
&
\!\beta_{20}\!=\!v_{xx};\\
 & \\
u_t=(u^{1/2}u_x)_x+f_{11}u^{-1/2}v+f_{12}u^{-1/2}+f_{13}u,
&
h_{21}=u_{xx}+{\displaystyle \frac{u_x^2}{2u}},\\
v_t=d_1v_{xx}+f_{21}v+f_{22}u^{3/2}+f_{23},
&
\beta_{21}=v_{xx};\\
 & \\
\end{array}
$$
$$
\begin{array}{ll}
u_t=(u^{1/2}u_x)_x+f_{11}u^{3/4}v^{1/4}+f_{12}u^{3/4}+f_{13}u,
&
h_{22}=u_{xx}- {\displaystyle \frac{3u_x^2}{4u}},\\
v_t=d_1(v^{1/2}v_x)_x+f_{21}u^{1/4}v^{3/4}+f_{22}v^{3/4}+f_{23}v,
&
\beta_{22}=v_{xx}-{\displaystyle \frac{3v_x^2}{4v}};\\
 & \\
u_t=(u^{1/2}u_x)_x+f_{11}u^{3/4}v^{3/2}+f_{12}u^{3/4}+f_{13}u,
&
h_{23}=u_{xx}-{\displaystyle \frac{3u_x^2}{4u}},\\
v_t=d_1(v^{1/2}v_x)_x+f_{21}u^{1/4}v^{-1/2}+f_{22}v^{-1/2}+f_{23}v,
&
\beta_{23}=v_{xx}+{\displaystyle \frac{v_x^2}{2v}};\\
 & \\
u_t=(u^{1/2}u_x)_x+f_{11}u^{-1/2}v^{1/4}+f_{12}u^{-1/2}+f_{13}u,
&
h_{24}=u_{xx}+{\displaystyle \frac{u_x^2}{2u}},\\
v_t=d_1(v^{1/2}v_x)_x+f_{21}u^{3/2}v^{3/4}+f_{22}v^{3/4}+f_{23}v,
&
\beta_{24}=v_{xx}-{\displaystyle \frac{3v_x^2}{4v}};\\
 & \\
u_t=(u^{1/2}u_x)_x+f_{11}u^{-1/2}v^{1/4}+f_{12}u^{-1/2}+f_{13}u,
&
h_{25}=u_{xx}+{\displaystyle \frac{u_x^2}{2u}},\\
v_t=d_1(v^{1/2}v_x)_x+f_{21}u^{1/4}v^{-1/2}+f_{22}v^{-1/2}+f_{23}v,
&
\beta_{25}=v_{xx}+{\displaystyle \frac{v_x^2}{2v}};\\
 & \\
u_t=(u^{1/2}u_x)_x+f_{11}u^{3/4}v+f_{12}u^{3/4}+f_{13}u,
&
h_{26}=u_{xx}-{\displaystyle \frac{3u_x^2}{4u}},\\
v_t=d_1(u^{1/2}v_x)_x+f_{21}v+f_{22}u^{1/4}+f_{23},
&
\beta_{26}=v_{xx};\\
 & \\
u_t=(v^{1/2}u_x)_x+f_{11}u+f_{12}v^{1/4}+f_{13},
&
h_{27}=u_{xx},\\
v_t=d_1(v^{1/2}v_x)_x+f_{21}uv^{3/4}+f_{22}v^{3/4}+f_{23}v,
&
\beta_{27}=v_{xx}-{\displaystyle \frac{3v_x^2}{4v}};\\
 & \\
u_t=(u^{1/2}u_x)_x+f_{11}u^{3/4}v+f_{12}u^{3/4}+f_{13}u,
&
h_{28}=u_{xx}-{\displaystyle \frac{3u_x^2}{4u}},\\
\!v_t\!=\!d_1(u^{1/4}v_x)_x\!+\!f_{21}v\!+\!f_{22}u^{1/4}\!+\!f_{23},
&
\!\beta_{28}\!=\!0;\\
 & \\
u_t=(v^{1/4}u_x)_x+f_{11}u+f_{12}v^{1/4}+f_{13},
&
h_{29}=u_{xx},\\
v_t=d_1(v^{1/2}v_x)_x+f_{21}uv^{3/4}+f_{22}v^{3/4}+f_{23}v,
&
\beta_{29}=v_{xx}-{\displaystyle \frac{3v_x^2}{4v}};\\
 & \\
u_t=(u^{1/2}u_x)_x+f_{11}u^{-1/2}v+f_{12}u^{-1/2}+f_{13}u,
&
h_{30}=u_{xx}+{\displaystyle \frac{u_x^2}{2u}},\\
v_t=d_1(u^{3/2}v_x)_x+f_{21}v+f_{22}u^{3/2}+f_{23},
&
\beta_{30}=v_{xx};\\
 & \\
\end{array}
$$
$$
\begin{array}{ll}
u_t=(v^{3/2}u_x)_x+f_{11}u+f_{12}v^{3/2}+f_{13},
&
h_{31}=u_{xx},\\
v_t=d_1(v^{1/2}v_x)_x+f_{21}uv^{-1/2}+f_{22}v^{-1/2}+f_{23}v,
&
\beta_{31}=v_{xx}+{\displaystyle \frac{v_x^2}{2v}};\\
 &  \\
u_t=(u^{1/2}u_x)_x+f_{11}u^{3/4}v^{1+a_1}+f_{12}u^{3/4}+f_{13}u,
&
h_{32}=u_{xx}-{\displaystyle \frac{3u_x^2}{4u}},\\
v_t=f_{21}u^{1/4}v^{-a_1}+f_{22}v^{-a_1}+f_{23}v,
&
\beta_{32}=v_{xx}+a_1{\displaystyle \frac{v_x^2}{v}};\\
 & \\
u_t\!=\!(u^{1/2}u_x)_x\!+\!f_{11}u^{-1/2}v^{1\!+\!a_1}\!+\!f_{12}u^{-1/2}\!
+\!f_{13}u,
&
h_{33}=u_{xx}+{\displaystyle \frac{u_x^2}{2u}},\\
v_t=f_{21}u^{3/2}v^{-a_1}+f_{22}v^{-a_1}+f_{23}v,
&
\beta_{33}=v_{xx}+a_1{\displaystyle \frac{v_x^2}{v}};\\
 & \\
u_t=(v^{1/2}u_x)_x+(3a_1v^{1/2}/4+f_{11})u+
&
h_{34}=u_{xx}+ {\displaystyle \frac{a_1}4}(u+f_{22}/f_{21}),\\
+ {\displaystyle \frac{3a_1f_{22}}{4f_{21}}}v^{1/2}+
{\displaystyle \frac{f_{11}f_{22}}{f_{21}}}\!),\\
v_t\!=\!f_{21}uv^{3/4}\!+\!f_{22}v^{3/4}\!+\!f_{23},
&
\beta_{34}\!=\!v_{xx}\!-{\displaystyle \frac{3v_x^2}{4v}}\!+a_1v.\\
\end{array}
$$

$$
\begin{array}{ll}
u_t=u_{xx}+f_1(u,v),
&
h_{35}=u_{xx}\!+\!f_1(u,v)\!+\!a_1u\!+\!a_2,\\
v_t\!=\!{\displaystyle \frac{a_1f_{1u}(u,v)u\!+\!a_2f_{1u}(u,v)\!-\!
a_1f_1(u,v)}{f_{1v}(u,v)}},
&
\beta_{35}=0;\\
 & \\
u_t=u_{xx}+uf_1(v)+a_1u\ln(u),
&
h_{36}=u_{xx}+uf_1(v)+a_1u\ln(u)+\\
 & +a_2e^{a_1t}u+a_3,\\
v_t=a_3 {\displaystyle \frac{f_1(v)+a_1(\ln(u)+1)}{uf'_1(v)}},
&
\beta_{36}=0;\\
& \\
u_t=u_{xx}+f_1\Bigg( {\displaystyle \frac{a_1v+a_2}{f_{21}e^{f_{23}u}}}\Bigg),
&
h_{37}=u_{xx}\!+\!a_1u_x\!+\!f_1\Bigg( {\displaystyle \frac{a_1v\!+\!
a_2}{f_{21}e^{f_{23}u}}}\Bigg)\!+\\
 & {\displaystyle \frac{a_3x+a_4}{f_{23}}},\\
v_t=f_{21}v+f_{22}e^{f_{23}u}+f_{24},
&
\!\beta_{37}\!=\!a_1v_x\!+\!f_{21}v\!+\!f_{22}
e^{f_{23}u}+\\
&\!+\!f_{24}\!+\!(a_3v\!+\!a_4)v\!+\! {\displaystyle \frac{f_{24}(a_3v\!+\!
a_4)}{f_{21}}};\\
\end{array}
$$
$$u_t=u_{xx}+f_{21}(a_2-1)u\ln(u)+uf_1\Bigg(\frac{v}{u^{a_1}}\Bigg),$$
$$v_t=f_{21}v+f_{22}u^{a_1},$$
$$h_{38}=u_{xx}+a_3e^{f_{21}(a_2-1)}u_x+f_{21}(a_2-1)u\ln(u)+uf_1
\Bigg(\frac{v}{u^{a_1}}\Bigg)+$$
$$+\frac{f_{21}a_3}{2}(a_2-1)\Bigg(e^{f_{21}(a_2-1)t}x+\frac1{a_4}\Bigg(a_5e^{f_
{21}(a_2-1)t}-\frac{2a_2a_4}{a_1a_3(a_2-1)}\Bigg)\Bigg)u,$$
$$\beta_{38}=a_3e^{f_{21}(a_2-1)t}v_x+f_{21}v+f_{22}u^{a_1}+\Bigg(\frac{a_1a_3
(a_2-1)f_{21}}{2a_4}(a_4x+a_5)e^{f_{21}(a_2-1)t}-f_{21}\Bigg)v;$$

$$u_t=u_{xx}+f_1(u)+\frac{a_1}{d_1}((n+1)v-d_1v^{n+1}),$$
$$v_t=d_1(v^nv_x)_x+\frac{d_1}{n+1}(v^{n+1}+a_2u+a_3)f_1'(u)-
\frac{a_2}{n+1}\Bigg((n+1-d_1n)v+\frac{d_1}{a_1}f_1(u)\Bigg),$$
$$h_{39}\!=\!u_{xx}\!+\!f_1(u)\!+\!\frac{a_1}{d_1}((n\!+
\!1)v\!-\!d_1v^{n\!+\!1})\!+\!a_2u\!+\!a_1v^{n\!+\!1}\!+\!a_3,$$
$$\beta_{39}=0;$$

$$u_t=u_{xx}-\frac{a_1}{d_1u}(d_1u^2v^{n+1}-(n+1)v)+f_1(u),$$
$$v_t=d_1(u^2v^nv_x)_x\!+\!\frac{d_1}{n\!+\!1}\Bigg(\Bigg(v^{n\!+\!1}\!+\!\frac{a_2}{a_1}\Bigg)
u^2\!+\!\frac{a_3}{a_1}u\Bigg)f_1'(u)\!-\!\frac{d_1}{n\!+\!1}\Bigg(v^{n\!+\!1}\!+$$
$$+\frac{a_2}{a_1}\Bigg)uf_1(u)-\frac{d_1a_3}{n+1}uv^{n+1}-2a_1v^{n+2}-2a_2v-
a_3\frac{v}{u},$$
$$h_{40}\!=\!u_{xx}\!-\!\frac{a_1}{d_1u}(d_1u^2v^{n\!+\!1}\!-\!(n\!+\!1)v)
\!+\!f_1(u)\!+\!(a_1v^{n\!+\!1}\!+\!a_2)u\!+
a_3,$$
$$\beta_{40}=0 ,$$
where $ f_{ij}$ and $a_k$ are arbitrary constants.

Moreover, we found all systems (4.1)-(4.2) which have solutions of the third order
$$h=u_{xxx}+h_1(t,x,u,v,u_1,v_1,u_2,v_2), $$
$$\beta=g_1v_{xxx}+\beta_1(t,x,u,v,u_1,v_1,u_2,v_2)$$
 to the linear determining equations (4.3)-(4.4).  The full list includes ten systems.
We give these systems together with solutions of the linear determining equations:
$$
\begin{array}{ll}
u_t=(uu_x)_x+f_{11}u^2+f_{12}u+f_{13},
&
h_{41}=u_{xxx}+f_{11}u_x/2,\\
v_t=d_1v_{xx}+f_{21}u+f_{22}v+f_{23},
&
\beta_{41}=v_{xxx}+f_{11}v_x/2;\\
 & \\
u_t=(vu_x)_x+(f_{11}v+f_{12})u+f_{13}v+f_{14},
&
h_{42}=u_{xxx}+f_{11}u_x/2,\\
v_t=d_1v_{xx}+f_{21}u+f_{22}v+f_{23},
&
\beta_{42}=v_{xxx}+f_{11}v_x/2;\\
 & \\
u_t=u_{xx}+f_{11}u+f_{12}v+f_{13},
&
h_{43}=u_{xxx}+f_{21}u_x/2,\\
v_t=d_1(vv_x)_x+d_1f_{21}v^2+f_{22}v+f_{23}u+f_{24}
&
\beta_{43}=v_{xxx}+f_{21}v_x/2; \\
 & \\
u_t=u_{xx}+f_{11}u+f_{12}v+f_{13},
&
h_{44}=u_{xxx}+f_{21}u_x/2, \\
v_t=d_1(uv_x)_x+(f_{21}u+f_{22})v+f_{23}u+f_{24},
  &
\beta_{44}=v_{xxx}+f_{21}
v_x/2;\\
 & \\
u_t=(uu_x)_x+f_{21}u^2+f_{12}u+f_{13}v+f_{14},
&
h_{45}=u_{xxx}+f_{21}u_x/2,\\
v_t=d_1(uv_x)_x+(d_1f_{21}u+f_{22})v+f_{23}u+f_{24},
&
\beta_{45}=v_{xxx}+f_{21}v_x/2;\\
 & \\
u_t=(vu_x)_x+(f_{21}v+f_{12})u+f_{13}u+f_{14},
&
h_{46}=u_{xxx}+f_{21}u_x/2,\\
v_t=d_1(vv_x)_x+d_1f_{21}v^2+f_{22}v+f_{23}u+f_{24},
&
\!\beta_{46}\!=\!v_{xxx}\!+\!f_
{21}v_x/2;\\
 & \\
u_t\!=\!(uu_x)_x\!+\!f_{21}u^2\!+\!f_{12}u\!+\!f_{13}v\!+\!f_{14},
&
h_{47}\!=\!u_{xxx}\!+\!f_{21}u_x/2,\\
v_t\!=\!d_1(vv_x)_x\!+\!d_1f_{21}v^2\!+\!f_{22}v\!+\!f_{23}u\!+\!f_{24},
&
\beta_{47}\!=\!v_{xxx}\!+\!f_{21}v_x/2;\\
 & \\
u_t\!=\!(vu_x)_x\!+\!(f_{21}u\!+\!f_{12})v\!+\!f_{13}u\!+\!f_{14},
&
h_{48}\!=\!u_{xxx}+f_{21}u_x/2,\\
v_t\!=\!d_1(uv_x)_x\!+\!(d_1f_{21}u\!+\!f_{22})v\!+\!f_{23}u\!+\!f_{24},
&
\beta_{48}=v_{xxx}+f_{21}v_x/2;\\
 & \\
u_t=(uu_x)_x+f_{11}u^2+f_{12}u+f_1(v),
&
h_{49}=u_{xxx}+f_{11}u_x/2,\\
v_t=(f_{21}u+f_{22}f_1(v)+f_{23})/f'_1(v),
&
\beta_{49}=v_{xxx}+3{\displaystyle \frac{f''_1(v)}{f'_1(v)}}v_{xx}v_x\!+\\
 & \!{\displaystyle \frac{f'''_1(v)}{f'_1(v)}}v_x^3\!+\!f_{11}v_x/2;\\
 & \\
u_t=(vu_x)_x\!+\!(f_{11}v\!+\!f_{12})u\!+\!f_{13}v\!+\!f_{14},
&
h_{50}\!=\!u_{xxx}\!+\!f_{11}u_x/2,\\
v_t\!=\!f_{21}u\!+\!f_{22}v\!+\!f_{23},
&
\beta_{50}\!=\!v_{xxx}\!+\!f_{11}v_x/2.
\end{array} $$

As some illustration of the basic method of computing solutions, we consider the system
$$u_t=(uu_x)_x+2u^2-3u+v,    \eqno{(4.5)}$$
$$v_t=(vv_x)_x+2v^2-2v+u,     \eqno{(4.6)}$$
with differential constraints
 $$u_{xxx}+u_x=0, \ \ v_{xxx}+v_x=0.  ,     \eqno(4.7)$$
It follows from (4.7) that the functions $u$ and $v$ have the representation
$$u=u_1(t)\sin(x)+u_2(t)\cos(x)+u_3(t), \
v=v_1(t)\sin(x)+v_2(t)\cos(x)+v_3(t). $$
Substituting this representation into the system (4.5)-(4.6), we obtain ordinary differential equations
$$u'_1=3u_1(u_3-1)+v_1,$$
$$u'_2=3u_2(u_3-1)+v_2,$$
$$u'_3=u_3(2u_3-3)+u_1^2+u_2^2+v_3,$$
$$v'_1=v_1(3v_3-2)+u_1,$$
$$v'_2=v_2(3v_3-2)+u_2,$$
$$v'_3=2v_3(v_3-1)+v_1^2+v_2^2+u_3.$$

\vspace{15mm}
 {\LARGE 5. Invariant solutions under involutive distributions.}

In this section we introduce invariant solutions under involutive
distributions.
%We consider the problem of finding involutive
%distributions that enable us to obtain invariant solutions to
%evolution equations.
Suppose that a collection of $p$ vector fields
$$X_{s}=\sum^{n}_{i=1} \xi^{i}_{s}(x)\partial_{x_{i}}$$
 is given on an open set $U\subset R^{n}$. If this collection is
 linearly disconnected, i.e., the rank of the matrix $|\xi^{i}_{s}(x)|$
 equals $p$ for all $x\in U$ and satisfies the involution condition
$$[X_{i},X_{j}]=\sum^{p}_{k=1}c^{k}_{ij}(x)X_{k}, \qquad
 \forall \ 1\leq i,j\leq p,     \eqno (5.1)$$
 where $c^{k}_{ij}$ are smooth functions, then this collection
generates an involutive p-dimensional distribution $D_p$. A
collection of vector fields with these properties is called an
involutive basis or just a basis. It is well known  that a
distribution $D_p$ is involutive if and only if it possesses at
least one involutive basis.

{\it Definition.} A solution $u = \varphi$  to a system of partial
differential equation $E$ is invariant under an involutive
distribution $D_p$ if $D_p$ is tangent to the manifold $S=\{(x,u):
\ u=\varphi(x) \}.$  Obviously, the invariance of a solution under
$D_p$ amounts to its invariance under the operators of an
arbitrary involutive basis for $D_p$.

Now, consider the system of evolution equations
 $$u^{i}_{t} = F^{i}(t,x,u,u_{\alpha}), \quad i=1,...,m, \eqno (5.2)$$
where $t$ and $x = (x_1,\dots,x_n)$ are independent variables,
$u^{1},...,u^{m}$ are functions, $u=(u^{1},...,u^{m})$, and
$u_{\alpha}$ stands for various partial derivatives with respect
to $x_1,\dots,x_n$. Denote the total derivatives with respect to
$t$ and $x_i$ by the symbols $D_t$ and $D_{x_i}$.

Let $J^{k}(U,R^{m})$ be the space of k-jets on $U\subset R^{n}$.
Recall that a manifold $H \subset J^{k}(R^{n+1},R^{m}),$ defined
by the equations
 $$h^{j}(t,x,u,u_{\beta})=0 , \quad j=1,...,s, \eqno (5.3)$$
is an invariant manifold for (5.2) if the following identity holds
on the set $[E] \bigcap [H]$:
 $$D_{t}h^{j}=0.$$
 Here $[E]$ and $[H]$ stand for the differential consequences of
(5.2) and (5.3) with respect to $x_{1},...,x_{n}$. Denote the
involutive distribution generated by vector fields
$X_{1},...,X_{r}$ by $<X_{1},...,X_{r}>.$

{\bf Lemma 1.} Suppose that vector fields $$X_{k}=\sum_{i=1}^{n}
\xi_{k}^{i}(t,x,u) \partial_{x_{i}} + \sum_{j=1}^{m}
\eta_{k}^{j}(t,x,u) \partial_{u^{j}} , \ k=1,...,n, \eqno (5.4)$$
generate an involutive distribution and that
$det(\xi_{k}^{i})\neq0.$ If the manifold defined by the equations
$$h_{k}^{j}=\sum_{i=1}^{n}\xi_{k}^{i}u_{x_{i}}^{j}-\eta_{k}^{j}=0,
\quad 1\leq j\leq m,  1\leq k \leq n,   \eqno (5.5)$$
 is invariant with respect to (5.2) then system (5.2) has invariant
solutions under this involutive distribution.

{\it Proof.} Write down the collection of the fields
$X_{1},...,X_{n}$ in vector form as follows:
 $$X=\xi\partial_{x} + \eta\partial_{u} .$$
 Acting by the matrix $\xi^{-1}$ on $X$, we obtain the involutive collection
$$Z=\partial_{x} + \tilde{\eta}\partial_{u} ,$$
 where $\tilde{\eta}=\xi^{-1}\eta.$ The distribution  $<Z_{1},...,Z_{n}>$ is
 involutive.
%as expressible in terms of an involutive distribution (see a
%proof, for example, in \cite[]).

The invariant solutions under $<X_{1},...,X_{n}>$ must satisfy
(5.5). The invariant solutions under $<Z_{1},...,Z_{n}>$  must
satisfy the equations
 $$u_{x_{k}}^{j}=\tilde{\eta}_{k}^{j}(t,x,u).\eqno (5.6)$$
Obviously, (5.5) and (5.6) have the same solutions. Since $Z$ is
an involution distribution, the Poisson bracket $[Z_{i},Z_{k}]$
vanishes.  Consequently, we have
 $$Z_{i}(\tilde{\eta}_{k}^{j})=Z_{k}(\tilde{\eta}_{i}^{j}),$$
which means that the consistency conditions for (5.6) are
satisfied.

Using (5.6) and inserting the derivatives of the functions $u^{j}$
with respect to $x_{k}$ in the right-hand side of (5.2), we come
to the system
 $$u_{t}^{j}=G^{j}(t,x,u), \qquad j=1,...,m.  \eqno (5.7)$$
 By the Frobenius theorem, the system of (5.6) and (5.7) is compatible if
the relations
 $$D_{x_{k}}G^{j}=D_{t}\tilde{\eta}_{k}^{j} , \qquad j=1,...,m;
\ k=1,...,n \eqno (5.8)$$
 are valid by virtue of (5.7) and (5.8).
Validity of these conditions follows from the invariance of (5.5)
with respect to (5.2). Indeed, this invariance means that
 $$D_{t} \left( u_{x_{k}}^{j} - \tilde{\eta}_{k}^{j} \right) =
 D_{x_k}F^{j}-D_{t}\tilde{\eta}_{k}^{j}=0. \eqno (5.9)$$
Inserting the derivatives with respect to $x_{k}$ in (5.9), we see
that (5.9) coincides with (5.8).

{\it Remark.} If an involutive distribution is generated by
analytic vector fields $X_{1},...,X_{p},$ where $p < n$, (5.2) is a
system of first-order equations with analytic right-hand sides,
and the rank of the matrix $\left(\xi_{k}^{i}\right)$ equals p,
then (5.2) has an invariant solution relative to
$X_{1},...,X_{p}$.  The proof is carried out by the above scheme,
but instead of the Frobenius theorem we should use the Riquier
theorem on the existence of analytic solutions to an autonomous
system with analytic right-hand sides \cite{pom}.

To exemplify the application of a distribution to constructing
solutions, consider the equation
 $$u_{t}=\Delta \ln u, \quad  \Delta=\frac{\partial^{2}}{\partial
x^{2}}+\frac{\partial^{2}}{\partial y^{2}},\eqno (5.10)$$
  which arises in various application \cite{ar,puk} and possesses an infinite-dimensional
algebra of point symmetries \cite{nar}. Some exact solutions to
this equation can be found in \cite{dor, gal}. We give a solution
to this equation which is invariant relative to the pair of
commuting operators
 $$X_{1}=\partial_{x} -(u^2 +(tu^2-xu^2+u) \tan(t))\partial_{u},$$
$$X_{2}=\partial_{y} -(tu^2+u-xu^2)\partial_{u} .$$
 The corresponding manifold for these vector fields is
 $$u_{x} + u^2 +(tu^2-xu^2+u) \tan(t) = 0, \eqno (5.11)$$
 $$u_{y} + tu^2+u-xu^2 = 0.         \eqno (5.12)$$
It is easy to verify that this is an invariant manifold for
(5.10). Note that the vector fields $X_{1}$ and $X_{2}$ do not
belong to the algebra of symmetries of (5.10).
 The general solution to (5.11),(5.12) and (5.10) has the form
 $$u=\frac{1}{A [\exp((x-t) \tan(t)+y] \cos t+x-t} , \quad A\in R .$$

To use vector fields and distributions, we need a method for
finding them. The classical approach to constructing vector fields
relative to which the given differential equations are invariant
was proposed by S. Lie. A modern exposition with many examples and
new results was given by L. V. Ovsyannikov \cite{ovs}.

A determining  equation like (2.5) enables us to find differential
constraints compatible with the original equation. In the case of
differential equations in more than two independent variables, we
can propose systems of defining equations which would enable us to
find involutive distributions.

Consider the system of involution equations (5.2) and the manifold
in $J^{1}(U,R^{m})$ defined by
 $$h_{j}^{i}=u_{x_{j}}^{i} + g_{j}^{i}(t,x,u)=0 , \eqno(5.13)$$
where $i=1,...,m,$ and $ j=1,...,n.$

{\it Theorem.} Suppose that the manifold (5.13) is invariant under
the system (5.2) whose right-hand sides are polynomials in
derivatives whose coefficients depend on $t,x_{1},...,x_{n}$ and
$u^{1},...,u^{m}.$ Then the functions  $h_{j}^{i}$ satisfy the
following system:

$$D_{t}h_{j}^{i} + m_{ij}(h) |_{[E]}=0, \ 1 \leq i \leq m, \ {1}
\leq {j} \leq {n}.\  \eqno (5.14)$$
 Here  $m_{ij}(h)$ is some operator representing a polynomial in
$h_{l}^{k},$ $ D_{x_{1}}h_{l}^{k},...,$ $
D_{x_{n}}h_{l}^{k},...,D^{\alpha}h_{l}^{k} $ $\ (k=1,...,m, \
l=1,...,n).$ The operators $m_{ij}(h)$ vanish whenever all
$h_{l}^{k}$ are zero.

{\it Proof.} We first show that the total derivative of $h_{j}^i$
with respect to $t$ is representable as
 $$D_{t}h_{j}^{i}=m_{ij}(h) + \gamma_{ij}, \eqno (5.15)$$
 where $m_{ij}$ are operators whose shape is described in the theorem
and $\gamma_{ij}$ are functions which may depend only on $t,x$ and
$u$.

The following identities are valid on $[E]$:
 $$D_{t}h_{j}^{i}=D_{x_{j}}F^{i} + \frac{\partial g_{j}^{i}}{\partial t} +
\sum_{k=1}^{m}F^{k} \frac{\partial g_{j}^{i}}{\partial u^{k}} .
\eqno (5.16)$$

Let  $\displaystyle\frac{\partial^{|s|} u^{k}}{\partial
x_{1}^{s_{1}} \cdots \partial x_{n}^{s_{n}}}$  be a derivative of
maximal order on the right-hand side of (5.16) and $s_p \neq 0$
for some $p$. By (5.16) and the assumptions of the theorem, this
derivative enters (5.16) polynomially. Using (5.13), we can write
down this derivative as follows:
 $$D_{x_{1}}^{s_{1}} \cdots D_{x_{p}}^{s_{p} - 1} \cdots
D_{x_{n}}^{s_{n}}(h_{p}^{k}) - D_{x_{1}}^{s_{1}} \cdots
D_{x_{p}}^{s_{p}-1} \cdots D_{x_{n}}^{s_{n}}(g_{p}^{k}).$$
 Note that the second summand involves no derivatives of order $|s|$ and is
a polynomial in derivatives. Thus, all derivatives of maximal
order on the right-hand side of (5.16) can be expressed in terms
of the total derivatives of the functions $h_{q}^{r}(r=1,...,m,$
and $q=1,...,n ).$ Afterwards, it is possible to express the
derivatives of order $|s| - 1$, etc. down to the first-order
derivatives.

 We are left with demonstrating that the functions $\gamma_{ij}$
in (5.15) are all zero. By the conditions of the theorem, the
manifold (5.13) is an invariant manifold for (5.2). Consequently,
the following identity holds on $[E] \bigcap [H]$:
 $$m_{ij}(h) + \gamma_{ij}=D_{t}h_{j}^{i}=0.$$
 Since the $m_{ij}$'s vanish on $[H]$, the functions $\gamma_{ij}$
are zero on $[E] \bigcap [H]$. Once the $m_{ij}$'s are independent
of the derivatives of the functions $u^{k}$, all $m_{ij}$ are
identically zero.

{\it Remark.} As we see from the proof of the theorem, the choice
of the operators $m_{ij}$ could be not uniquely defined.

For example, consider the second-order equation in three
independent variables:
 $$u_{t} = G \equiv F^1 u_{xx} + F^2 u_{yy} + F^3 u_{x}^{2} +
F^4 u_{y}^2 + F^{5} , \eqno( 5.17)$$
 where $F^i$ are some functions depending on $u$. Suppose that
 $$h_{1}\equiv u_{x} + g_{1}(t,x,y,u)=0, \
h_{2}\equiv u_{y} + g_{2}(t,x,y,u)=0 \eqno(5.18)$$
 define an invariant manifold for (5.17). To derive a system of
determining equations like (5.14), we express the derivatives
$D_{t}h_{1}$ and $D_{t}h_{2}$ in terms of $h_{i},D_{x}h_{i}$,
$D_{y}h_{i}$, $D_{x}^{2}h_{i},D_{x}D_{y}h_{i}$, and
$D_{y}^{2}h_{i}$ ($i = 1,2$).   By (5.17), the following holds:
$$D_{t}h_{1}=D_{x}G + \frac{\partial g_{1}}{\partial t} +
\frac{\partial g_{1}}{\partial u}G . $$
 It is easy to verify that
the right-hand side of the last equality is representable as
 $$m_{11}(h_{1},h_2) = G_{u_{xx}}D_{x}^{2}h_{1} + G_{u_{yy}}D_{y}^{2}h_{1}
 + [G_{u_{x}} + D_{x}(G_{u_{xx}})]D_{x}h_{1} + G_{u_{y}}D_{y}h_{1} +$$
$$ + D_{x}(G_{u_{yy}})D_{y}h_{2} + [ G_{u} - D_{x}^{2}(G_{u_{xx}})
- D_{y}^{2}(G_{u_{yy}})+ r_{1} ]h_{1}
 + s_{1}h_{2} + \gamma_{1},  \eqno(5.19)$$
 where $r_{1},s_{1},$ and $\gamma_{1}$ are functions depending on
 $h_1, h_{2},$ and $G$. Since (5.18) is an invariant manifold, the
function $\gamma_{1}$ equals 0. Consequently, the first defining
equation has the form
 $$D_{t}h_{1} = m_{11}(h_{1},h_{2}).$$
 To obtain the second defining equation
 $$D_{t}h_{2} = m_{12}(h_{1},h_{2}) ,$$
 we should replace $h_{1}$ in (5.12) with  $h_{2}$, $x$ with $y$,
$r_{1}$ with $r_{2}$, and $s_{1}$ with $s_{2}$.
 The following lemma asserts that, under some conditions,
 solutions to equations like (3.13) enable us
to construct differential constraints compatible with the system
of evolution equations (5.2). It is worth to note that the form of
the operators $m_{ij}$ is inessential, provided that only
$m_{ij}(0)=0.$

{\bf Lemma 2. } Suppose that the functions
$$h_{j}^{i}=\sum_{s=1}^{n}\xi_{j}^{s}(t,x,u)u_{x_{s}}^{i} -
g_{j}^{i}(t,x,u)$$
 satisfy a system like (5.14) on $[E]$ with $m_{ij}(0)=0.$
If the vector fields
 $$X_{j}=\sum_{s=1}^{n}\xi_{j}^{s}\partial_{x_{s}} +
\sum_{i=1}^{m}g_{j}^{i}\partial_{u_{i}}, \ j=1,...,n$$
 generate an involutive distribution and $\det(\xi_{j}^{s}) \neq 0$
then there is a solution to the system of (5.2) and the equations
$$h_{j}^{i}=0 , \qquad  i=1,...,m, \ j=1,...,n . \eqno(5.20)$$
 {\it Proof.} Since the functions  $h_{j}^{i}$
satisfy (5.14), in view of $m_{ij}(0)= 0$  (5.20) defines an
invariant manifold for (5.2). To complete the proof, it suffices
to refer to Lemma 1.

Finding solutions to general nonlinear equations (5.14) might
represent a very complicated problem. To simplify the problem, we
remove all terms nonlinear in $h_{l}^{k}$ from the operators
$m_{ij}$ as it was done above in the case of an evolution equation
with one space variable. In result, we obtain some linear equation
 $$D_{t}h_{j}^{i} + l_{ij}(h)=0. $$
As we done above, multiply the coefficients of the operators
$l_{ij}$ by undetermined constants and write down the resultant
equations as
 $$D_{t}h_{j}^{i} + L_{ij}(h)=0 \eqno(5.21)$$
 calling them linear determining equations (LDEs).
 For example, the LDEs for (5.17) have the form
 $$D_{t}h_{1} = L_{11}(h_{1},h_{2})\equiv a_{1}G_{u_{xx}}D_{x}^{2}h_{1} +
a_{2}G_{u_{yy}}D_{y}^{2}h_{1} +$$ $$+ [a_{3}G_{u_{x}}
+a_{4}D_{x}(G_{u_{xx}})]D_{x}h_1
 +a_{5}G_{u_{y}}D_{y}h_{1} + a_{6}D_{x}(G_{u_{yy}})D_{y}h_{2} +$$
$$+[a_{7}G_{u} + a_{8}D_{x}^{2}(G_{u_{xx}})
 + a_{9}D_{y}^{2}(G_{u_{yy}})]h_{1}, \eqno(5.22)$$
$$D_{t}h_{2} = L_{12}(h_{1},h_{2}),$$
 where $L_{12}(h_{1},h_{2})$ is obtained from $L_{11}(h_{1},h_{2})$
by replacing $h_{1}$ with $h_{2}$, $x$ with $y$, and $a_{i}$ with
$b_{i}$.

Although the above arguments were for systems of evolution
equations, we can try to extend them to a more general situation.
Assume given a system
 $$n_{i}(u) = F^{i}(t,x,u,u_{\alpha}),\qquad i=1,...,m,$$
where $n_{i}$ are linear differential operators with constant
coefficients and the right-hand sides are similar to those in the
case of evolution systems (5.2). To find the functions
$h_{j}^{i}$, we suggest using the following equation in place of
(5.21):
 $$N_{i}(h_{j}^{i}) + L_{ij}(h)=0, \eqno(5.23)$$
where the operators  $N_{i}$ are obtained from $n_{i}$  by
replacing partial derivatives with total derivatives. Alongside
(5.23), it is useful to introduce the following analog of
B-defining equations \cite{kap3}:
 $$N_{i}(h_{j}^{i}) + L_{ij}(h) +
\sum_{1\leq l\leq m \atop 1\leq k\leq n}b_{lj}^{ki}h_{k}^{l}=0,
\eqno(5.24)$$
 where $1 \leq i \leq m, 1 \leq j \leq n,$ and $b_{lj}^{ki}$
are functions that may depend on $t, x,$ and $u$.

We call equations of the form (3.23) quasilinear determining
equations (QDEs). We exhibit an example of QDEs in finding
involutive distributions. Consider one of the nonlinear dispersion
models describing the propagation of long two-dimensional waves
[22]:
 $$\eta_{tt}=gd\Delta\eta + \frac{d^{2}}{3}\Delta\eta_{tt} +
\frac{3}{2}g\Delta\eta^{2},$$
 where $\eta(t,x,y)$ is the deviation
of a fluid from an equilibrium state, $d$ is the depth of an
unperturbed fluid, and $g$ is the free fall acceleration. By
translations and dilations, we can reduce this equation to the
form
 $$u_{tt} - \Delta(u_{tt}) - u\triangle u - (\nabla u)^{2}=0. \eqno(5.25)$$
In accordance with the above method, the QLEs for (5.25) have the
form
 $$D_{t}^{2}h_{1} - D_{t}^{2}D_{x}^{2}h_{1} -
D_{t}^{2}D_{y}^{2}h_{1} + a_{1}u(D_{x}^{2}h_{1} + D_{y}^{2}h_{1})
+ a_{2}u_{x}D_{x}h_{1} + a_{3}u_{y}D_{y}h_{1}$$ $$+
a_{4}u_{x}D_{y}h_{2} + (a_{5}\Delta u + a_{6}u_{xx} + a_{7}u_{yy}
+ r_{1})h_{1} + q_{1}h_{2}=0,   \eqno(5.26)$$
 $$D_{t}^{2}h_{2} - D_{t}^{2}D_{x}^{2}h_{2} -
D_{t}^{2}D_{y}^{2}h_{2} + b_{1}u(D_{x}^{2}h_{2} + D_{y}^{2}h_{2})
+ b_{2}u_{y}D_{y}h_{2} + b_{3}u_{x}D_{x}h_{2}$$ $$+
b_{4}u_{y}D_{x}h_{1} + (b_{5}\Delta u + b_{6}u_{xx} + b_{7}u_{yy}
+ r_{2})h_{2} + q_{2}h_{1} =0,  \eqno(5.27)$$
 where $a_{i}$ and
$b_{i}$ are constants, and $r_{j}$ and $q_{j}$ are functions which
may depend on $t, x, y,$ and $u$ and which should be found
together with $h_{1}$ and $h_{2}.$ The scheme for solving (5.26)
and (5.27) is completely analogous to the standard scheme of the
group analysis of differential equations \cite{ovs,olv}. For this
reason, we omit all intermediate computations and set forth only
the final results.

If $h_{1}$ and $h_{2}$ are sought in the form corresponding to the
point symmetries
 $$h_{1}=\xi_{1}^{1}u_{t} + \xi_{2}^{1}u_{x} +
\xi_{3}^{1}u_{y} + \eta^{1},$$ $$h_{2}=\xi_{1}^{2}u_{t} +
\xi_{2}^{2}u_{x} + \xi_{3}^{2}u_{y} + \eta^{2},$$
 where $\xi^{i}$ and $\eta^{j}$  are functions of $t, x, y,$
and $u$, then under the condition
 $(\xi_{1}^{1})^{2} + (\xi_{3}^{1})^{2} + (\xi_{1}^{2})^{2}
+ (\xi_{2}^{2})^{2}\neq 0$
 equations (5.26) and (5.27) can be shown to have
solutions leading only to admissible operators for (5.25). There
appear new solutions only when
 $$h_{1}=u_{x} + g_{1}(t,x,y,u),\ h_{2}=u_{y} +g_{2}(t,x,y,u). $$
The final form of $g_{1}$ and $g_{2}$ is as follows:
 $$g_{1}=s_{1}x + s_{2}y + s_{3},\ g_{2}=s_{2}x + s_{4}y + s_{5} .$$
Moreover, the functions $s_{i}$ $(i = 1,...,5)$ depend only on $t$
and satisfy the following system of five second-order differential
equations:
 $$s_{1}^{''} + 3s_{1}^{2} + s_{1}s_{4} + 2s_{2}^{2}=0,$$
$$s_{2}^{''} + 3s_{1}s_{2} + 3s_{2}s_{4}=0,$$ $$s_{3}^{''} +
3s_{1}s_{3} + 2s_{2}s_{5} + s_{3}s_{4}=0,$$ $$s_{4}^{''} +
s_{1}s_{4} + 2s_{2}^{2} + 3s_{4}^{2}=0,$$ $$s_{5}^{''} +s_{1}s_{5}
+ 2s_{2}s_{3} +3s_{4}s_{5}=0.$$
 For completeness of exposition, we
write down the constants $a_{i}$ and $b_{i}$ $(i=1,...,7)$ and the
functions $r_{j}$ and $q_{j}$ $(j=1,2)$ in (3.26) and (3.27)
corresponding to $g_{1}$ and $g_{2}$:
$$a_{1}=b_{1}=a_{4}=b_{4}=-1, \ a_{2} = b_{2} = a_{3} = b_{3} =
-3,$$ $$a_{5}=a_{6}=a_{7}=b_{5}=b_{6}=b_{7}=0,$$ $$r_{1}=3s_{1} +
s_{4},\ r_{2}=s_{1} + 3s_{4}, \ q_{1}=2s_{1},\ q_{2}=2s_{2}.$$
 The functions $h_1$ and $h_2$ generate the differential constraints
 $$u_{x} + s_{1}x + s_{2}y + s_{3}=0,$$
 $$u_{y} + s_{2}x + s_{4}y + s_{5}=0.$$
These constraints enable us to find the following representation
for a solution to (5.25):
  $$u=\frac{-s_{1}x^{2}}{2} - s_{2}xy -
\frac{s_{4}y^{2}}{2} - s_{3}x - s_{5}y + s_{6}.$$
 Inserting this in (5.25), we obtain the following equation for $s_6$:
$$s_{6}^{''} =3s_{1}^{2} + 2s_{1}s_{4} - s_{1}s_{6} + 4s_{2}^{2} +
s_{3}^{2} +3s_{4}^{2} - s_{4}s_{6} + s_{5}^{2}.$$
 The system of
the six differential equations in the six functions $s_i$ deserves
further study. For example, it would be interesting to find a
solution expressible via elementary functions.

%{\bf 6. Conclusion.}

\vspace{15mm} {\bf  Acknowledgement }

This work was supported in part by RFBR, project 01 - 01 - 00850.

\end{document}